\documentclass [useAMS]{mn2e}
\usepackage {graphicx}

\begin{document}

\title{Magnetic connection and current distribution in black hole accretion
discs}
\author[]{Cheng-Xuan Zhao$^1$,Ding-Xiong Wang$^{1,2}$ and Zhao-Ming Gan$^1$ \\
$1$ School of Physics, Huazhong University of Science and Technology, Wuhan,430074,China \\
$2$ Send offprint requests to: D.-X. Wang (dxwang@mail.hust.edu.cn)}
\maketitle

\begin{abstract}

We discuss one of the possible origins of large-scale magnetic
fields based on a continuous distribution of toroidal electric
current flowing in the inner region of the disc around a Kerr black
hole (BH) in the framework of general relativity. It turns out that
four types of configuration of the magnetic connection (MC) are
generated, i.e., MC of the BH with the remote astrophysical load
(MCHL), MC of the BH with the disc (MCHD), MC of the plunging region
with the disc (MCPD) and MC of the inner and outer disc regions
(MCDD). It turns out that the Blandford-Znajek (BZ) process can be
regarded as one type of MC, i.e., MCHL. In addition, we propose a
scenario for fitting the quasi-periodic oscillations in BH binaries
based on MCDD associated with the magnetic reconnection.
\end{abstract}


\section{INTRODUCTION}

According to observations, large-scale magnetic fields exist in many
astronomical cases, such as galaxies, young stellar objects, neutron stars
and black hole (BH) X-ray binaries (Han {\&} Qiao 1994; Livio Ogilvie {\&}
Pringle 1999). It is widely believed that magnetic fields play an important
role in dynamics of accretion discs and jet formation. However, the origin
of large-scale magnetic fields remains elusive (Blandford 2002).

Recently, much attention has been paid to the magnetic connection (MC) of a
rotating black hole (BH) with its surrounding accretion disc (Blandford
1999; Li 2000; Wang, Xiao {\&} Lei 2002; Uzdensky 2004, 2005). Not long ago,
Li (2002) calculated the magnetic field configuration of MC by assuming a
single electric current flowing in equatorial plane of a BH. Wang et al$.$
(2007) discussed the MC between plunging region and disc (MCPD) as well as
the MC between the BH horizon and the disc (MCHD). Very recently, Ge et al.
(2008) analyzed the topology and feature of a magnetic field configuration
arising from double counter oriented electric current-rings in an accretion
disc around a Kerr BH.

Enlightened by the above works, in this paper, we intend to investigate the
magnetic field configuration arising from toroidal electric current
distributed continuously in a thin disc around a Kerr BH, and discuss the MC
of the BH with its surrounding disc. In addition, we propose a scenario for
fitting the quasi-periodic oscillations in BH binaries based on the MC
associated with magnetic reconnection in the disc.

Throughout this paper the geometric units $G = c = 1$ are used.

\section{ MAGNETIC FIELD CONFIGURATION AND MAGNETIC FLUX }

In this paper, the magnetic field configuration is produced by the toroidal
electric current flowing in the inner region of an accretion disc based on
the following assumptions.

(1) The disc is thin and Keplerian around a Kerr BH, and the
concerned Kerr metric parameters are given by

\begin{equation}
\label{eq1}
\begin{array}{l}
 \rho ^2 = r^2 + a^2\cos ^2\theta ,\mbox{ }\Delta = r^2 - 2Mr + a^2, \\
 A = \left( {r^2 + a^2} \right)^2 - a^2\Delta \sin ^2\theta ,\\ \alpha =
 \left( {\rho ^2\Delta / A} \right)^{1 / 2},\mbox{ }\varpi
= \left( {A / \rho ^2} \right)^{1 / 2}\sin \theta . \\
 \end{array}
\end{equation}

(2) The toroidal electric current lies in the equatorial plane,
varying continuously with the disc radius in a power-law,

\begin{equation}
\label{eq2}
j = j_0 \left( {r \mathord{\left/ {\vphantom {r {r_{ms} }}} \right.
\kern-\nulldelimiterspace} {r_{ms} }} \right)^{ - n}\delta \left( {cos\theta
} \right),
\quad
r_{ms} \le r \le \lambda r_{ms} ,
\end{equation}

\noindent
where $n$ is the power-law index for the variation of the electric current,
$r_{ms} $ is the innermost stable circular orbit (ISCO), and $\lambda $ is a
parameter to adjust the radial width of the current distribution.

Thus the four-current vector flowing on the circle of radius ${r}'$ is given
by

\begin{equation}
\label{eq3} J^\alpha = \frac{1}{r}\left( {\frac{\Delta }{A}}
\right)^{1 \mathord{\left/ {\vphantom {1 2}} \right.
\kern-\nulldelimiterspace} 2} \left( {\frac{\partial }{\partial
\varphi }} \right)^\alpha \left. j \right|_{r = {r}'} .
\end{equation}

Based on the above distribution of the toroidal electric current we have the
magnetic flux through a surface bounded by a circle with $r = $\textit{const} and $\theta
= $\textit{const} as follows,

\begin{equation}
\label{eq4}
\Psi \left( {a_\ast ,n,\lambda ,r,\theta ,} \right) = 2\pi \cdot \int
{dA_\varphi \left( {a_\ast ,n,\lambda ,r,\theta ;{r}'} \right)} .
\end{equation}

In equation (\ref{eq4}) $dA_\varphi \left( {a_\ast ,n,\lambda
,r,\theta ;{r}'} \right)$ is the toroidal component of the electric
vector potential, which can be determined by the electric current
loop located at ${r}' - {r}' + d{r}'$ (Znajek 1978; Linet 1979), and
the position $(r,\theta )$ of the circle is described in the
spherical coordinates.

A mapping relation between two circles with spherical coordinates $(r_1
,\theta _1 )$ and $(r_2 ,\theta _2 )$ is given based on the conservation of
magnetic flux as follows,

\begin{equation}
\label{eq5}
\tilde {\Psi }\left( {a_\ast ,n,\lambda ,\tilde {r}_1 ,\theta _1 } \right) =
\tilde {\Psi }\left( {a_\ast ,n,\lambda ,\tilde {r}_2 ,\theta _2 } \right).
\end{equation}

In equation (\ref{eq5}) $\tilde {\Psi }$ is the dimensionless
magnetic flux, being defined as

\begin{equation}
\label{eq6}
\tilde {\Psi }\left( {a_\ast ,n,\lambda ,\tilde {r},\theta } \right) = {2\pi
A_\varphi \left( {a_\ast ,n,\lambda ,\tilde {r},\theta } \right)}
\mathord{\left/ {\vphantom {{2\pi A_\varphi \left( {a_\ast ,n,\lambda
,\tilde {r},\theta } \right)} {B_0 M^2}}} \right. \kern-\nulldelimiterspace}
{B_0 M^2},
\end{equation}

\noindent where we have $\tilde {r} \equiv r / M$ and $B_0 \equiv
2j_0 $. Based on (\ref{eq6}) we have the relation between the radii
$r_{ms} $ and $r_0 $ as follows,

\begin{equation}
\label{eq7}
\tilde {\Psi }\left( {a_\ast ,n,\lambda ,\tilde {r}_0 ,\pi \mathord{\left/
{\vphantom {\pi 2}} \right. \kern-\nulldelimiterspace} 2} \right) = \tilde
{\Psi }\left( {a_\ast ,n,\lambda ,\tilde {r}_{ms} ,\pi \mathord{\left/
{\vphantom {\pi 2}} \right. \kern-\nulldelimiterspace} 2} \right).
\end{equation}

Incorporating equations (\ref{eq1})---(\ref{eq7}), we have the
poloidal magnetic field configuration generated by continuous
toroidal electric current with different values of the parameters
$a_ * $, $n$ and $\lambda $ as shown in Fig.1.

\begin{figure}
\vspace{0.5cm}
\begin{center}
 {\includegraphics[width=5.6cm]{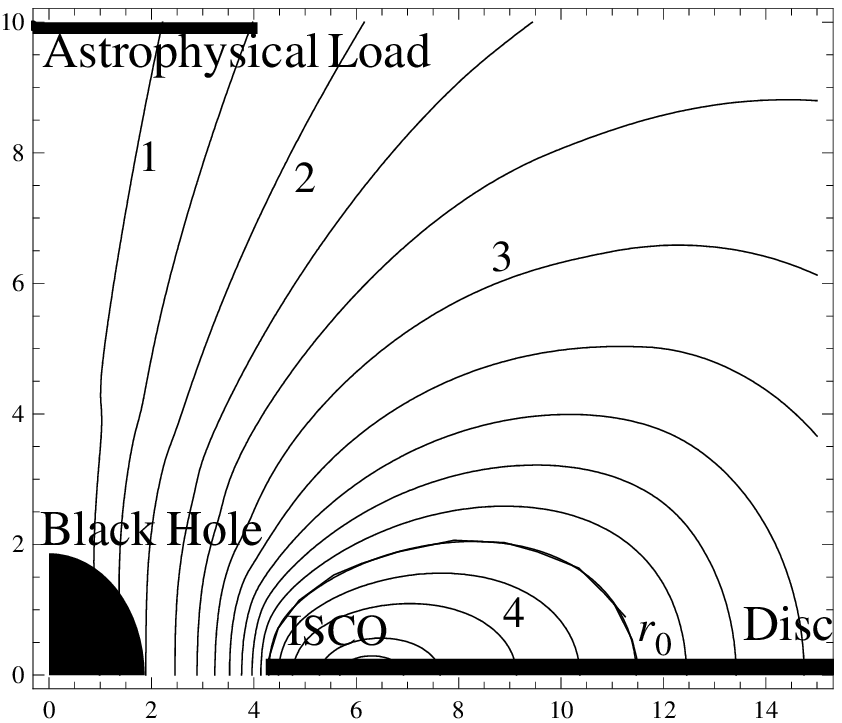}
   \centerline{\hspace{0.4cm}(a)}\\
  \includegraphics[width=5.6cm]{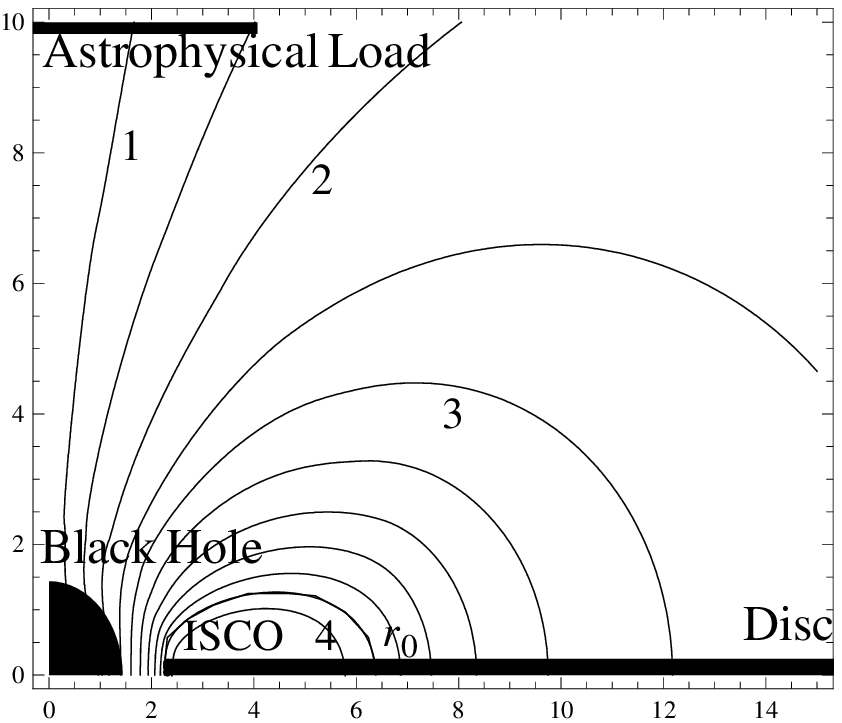}
   \centerline{\hspace{0.4cm}(b)}\\
  \includegraphics[width=5.6cm]{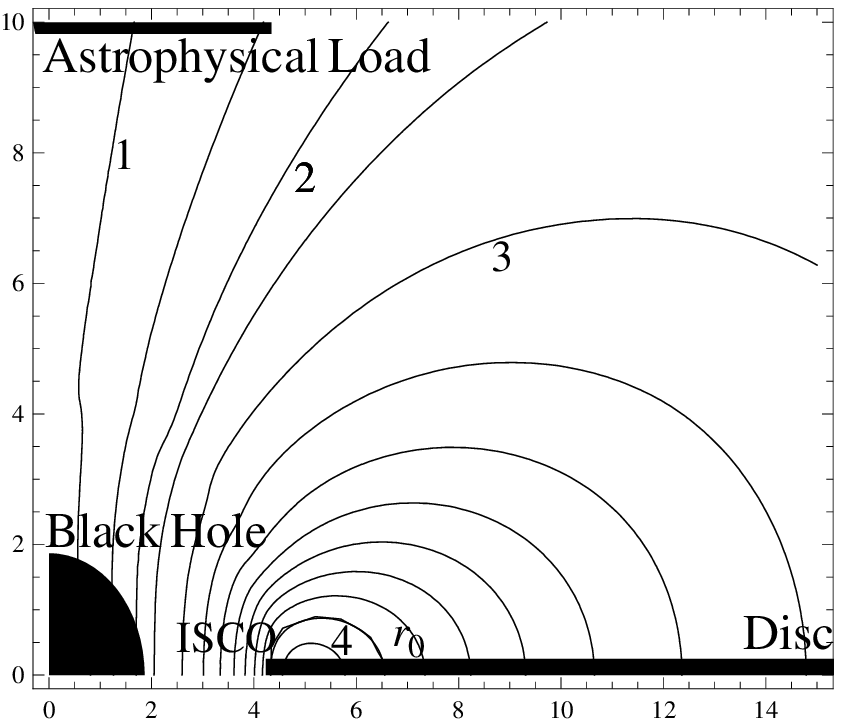}
   \centerline{\hspace{0.4cm}(c)}\\
  \includegraphics[width=5.6cm]{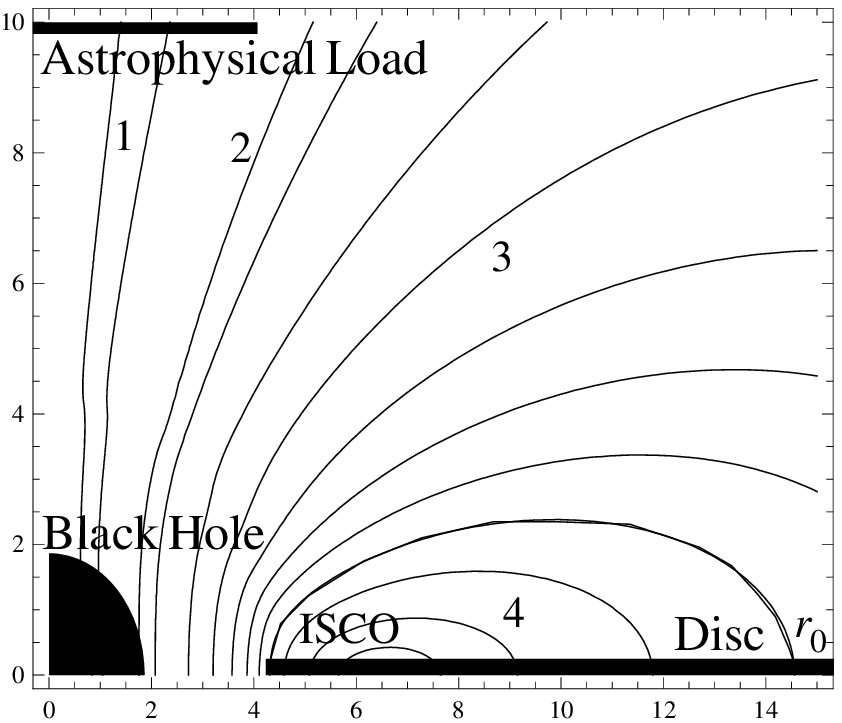}
   \centerline{\hspace{0.4cm}(d)}}

\caption{The magnetic field configuration generated by toroidal
electric current distributed continuously over the inner region of a
thin disc around a Kerr BH with (a) $a_*=0.5,n=3.0,\lambda=3$, (b)
$a_*=0.9,n=3.0,\lambda=3$, (c) $a_*=0.5,n=5.0,\lambda=3$ and (d)
$a_*=0.5,n=3.0,\lambda=5$. } \label{fig1}
\end{center}
\end{figure}

As shown in Fig.1, the four types of MC configurations can coexist
for different values of $a_\ast $, $n$ and $\lambda $, where the
characteristic field lines of Types I, II, III and IV are indicated
by lines 1, 2, 3 and 4, respectively. These MC configurations are
described as follows.

Type I: MC of the BH with the remote astrophysical load (MCHL);

Type II: MC of the BH with the disc (MCHD);

Type III: MC of the plunging region with the disc (MCPD);

Type IV: MC of the inner and outer regions of the disc (MCDD).

Among the above configurations MCHD and MCPD have been discussed by a number
of authors (Blandford 1999; Krolik 1999; Gammie 1999; Li 2002; Wang et al.
2002; Wang et al. 2007), and these two MC processes are regarded as the
variants of the Blandford-Znajek (BZ) process (Blandford {\&} Znajek 1977).
On the other hand, the BZ process can be also regarded as one kind of MC,
i.e., MCHL, which was formulated with the MC process in a united model
(Wang, Xiao {\&} Lei 2002).

Based on equation (\ref{eq7}) we have the magnetic flux $\left.
{\tilde {\psi }(r,\theta )} \right|_{\theta = \pi \mathord{\left/
{\vphantom {\pi 2}} \right. \kern-\nulldelimiterspace} 2} $
penetrating the equatorial plane of a Kerr BH versus the disc radius
$\tilde {r}$ for different power-law index $n$ as shown in Fig.2.

\begin{figure}
\begin{center}
{\includegraphics[width=6cm]{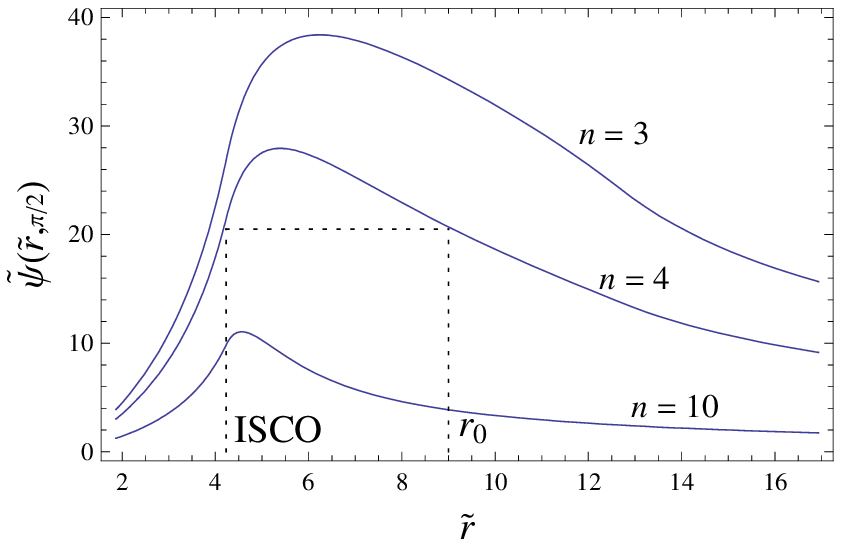}
 \centerline{\hspace{1cm}(a)}
 \includegraphics[width=6cm]{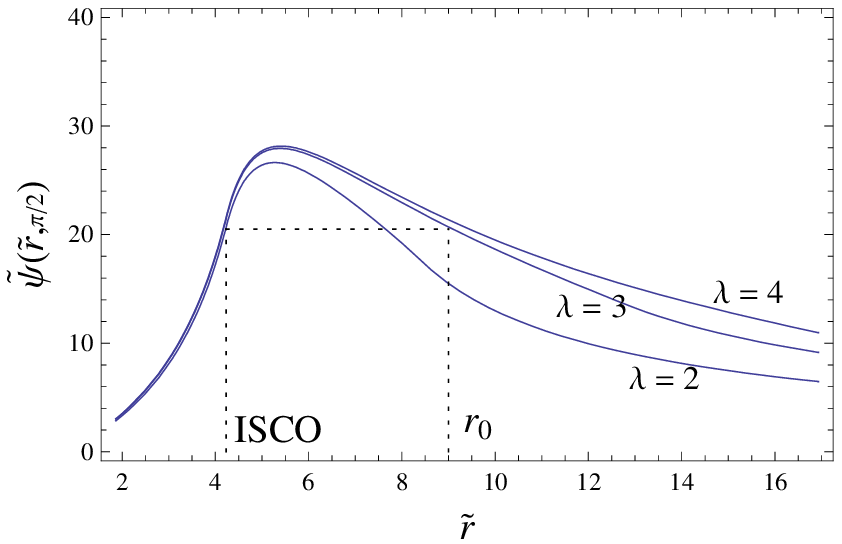}
 \centerline{\hspace{1cm}(b)}}
 \label{fig2}
 \caption{Magnetic
flux penetrating the equatorial plane of a Kerr BH versus $ \tilde
r$ for $a_*=0.5$ (a) $\lambda=3$ with $n=3,4$ and $10$, and (b)
$n=4$ with $\lambda=2,3$ and $4$.}
\end{center}
\end{figure}

From Fig.2 we obtain the following results.

(1) The magnetic flux $\left. {\tilde {\psi }(r,\theta )}
\right|_{\theta = \pi \mathord{\left/ {\vphantom {\pi 2}} \right.
\kern-\nulldelimiterspace} 2} $ varies with $\tilde {r}$
non-monotonically, and attains its peak value outside ISCO. The
greater the power-law index $n$ is, the less the peak value is, and
the closer the position of the peak value is to ISCO.

(2) Equation (\ref{eq7}) provides a relation between the radii
$r_{ms} $ and $r_0 $. For the given $a_ * $, the greater $n$
corresponds to the less $r_0 $, and the greater $\lambda $
corresponds to the greater $r_0 $ as shown in Figures 2a and 2b,
respectively.


\section{ FITTING QPOS OF BH BINARIES }

As is well known, kHz Quasi-Periodic Oscillations (QPOs) have been observed
in several X-ray Binaries. As argued by van der Klis (2000), kHz QPOs in
X-ray binaries probably originate from the inner edge of an accretion disc
with a BH of stellar-mass, since millisecond is the natural timescale for
accretion process in these regions. Although a number of models have been
proposed to explain the QPOs in X-ray Binaries, no consensus has been
reached on their physical origin (see a review by McClintock {\&} Remillard
2006, hereafter MR06).

\subsection{ Fitting QPO frequencies of BH X-ray binaries based on MCDD }

Now we propose a scenario for fitting QPOs in BH binaries based on
MCDD corresponding to the closed field lines indicated by line 4 in
Fig.1. The field lines of MCDD are frozen in the disc, and the
critical closed field line of MCDD connects the footpoint at ISCO
and that at radius $r_0 $ as shown by the thick solid line in Fig.1.

The Keplerian angular velocity depends on the BH mass and spin, and
decreases monotonically with the increasing disc radius, and it reads (Wang
et al. 2003)

\begin{equation}
\label{eq8}
\Omega _K = 2\pi \nu _0 (\xi ^{3 \mathord{\left/ {\vphantom {3 2}} \right.
\kern-\nulldelimiterspace} 2}\chi _{ms}^3 + a_\ast )^{ - 1},
\end{equation}

\noindent
where $\nu _0 $ is defined as $\nu _0 \equiv (m_{BH} )^{ - 1}3.23\times
10^4Hz$ with $m_{BH} \equiv M \mathord{\left/ {\vphantom {M {M_ \odot }}}
\right. \kern-\nulldelimiterspace} {M_ \odot }$, and parameter $\xi \equiv r
\mathord{\left/ {\vphantom {r {r_{ms} }}} \right. \kern-\nulldelimiterspace}
{r_{ms} }$ is the disc radius in terms of $r_{ms} $, and $\chi _{ms} \equiv
\sqrt {{r_{ms} } \mathord{\left/ {\vphantom {{r_{ms} } M}} \right.
\kern-\nulldelimiterspace} M} $ is a function of the BH spin $a_ * $
(Novikov {\&} Thorne 1973).

Since the inner footpoints of the magnetic field lines rotate faster than
the outer ones, the rotating disc will twist the field lines of MCDD,
resulting in an increasing toroidal component of the magnetic field. This
configuration allows antiparallel segments of the field lines to be brought
in contact, giving rise to magnetic reconnection in a similar way to that
between a central star and the accretion disc (Montmerle et al. 2000). The
magnetic energy thus librated could heat the plasma and ignites a flare. In
this way, successive reconnections and flaring may continue periodically for
some time, providing a possible way for producing QPOs in BH accretion disc.

Considering that the magnetic reconnection arises from the twist of the
closed field lines of MCDD, we infer that the flare occurs most probably in
the critical field line, and the difference between the Keplerian
frequencies at radii $r_{ms} $ and $r_0 $ as follows,

\begin{equation}
\label{eq9} \begin{array}{l}
 \nu _{MCDD} = \frac{\Omega _K \left(
{r_{ms} } \right) - \Omega _K \left( {r_0 } \right)}{2\pi } \\
\\ \quad\quad\quad=\nu _0 \left[ {\left( {\chi _{ms}^3 + a_\ast }
\right)^{ - 1} - \left( {\xi _0^{3 \mathord{\left/ {\vphantom {3 2}}
\right. \kern-\nulldelimiterspace} 2} \chi _{ms}^3 + a_\ast }
\right)^{ - 1}} \right].
\end{array}
\end{equation}

Thus we can fit the frequency $\nu _{QPO} $ of single--component
QPOs as $\nu _{MCDD} $ based on equation (\ref{eq9}), from which we
find that the frequency $\nu _{QPO} $ depends on three parameters,
$m_{BH} $, $a_\ast $ and $\xi _0 \equiv {r_0 } \mathord{\left/
{\vphantom {{r_0 } {r_{ms} }}} \right. \kern-\nulldelimiterspace}
{r_{ms} }$. By using equation (\ref{eq9}) we have an isosurface of
$\nu _{QPO} $=\textit{ const} in a parameter space consisting of
these parameters. Taking the BH binary GRO J1655-40 as an example,
we have an isosurface of $\nu _{QPO} = 300Hz$ with $\lambda = 3$ as
shown in Fig.3. The parameter $\xi _0 $ approaches the maximum 1.554
at point A with $m_{BH} = 6.6$ and $a_\ast = 0.65$, and the minimum
1.338 at point B with $m_{BH} = 6.0$ and $a_\ast = 0.8$. Thus we
have $n = 5.284$ and 6.656 by substituting $\xi _0 $= 1.554 and
1.338 respectively into (\ref{eq7}).

We also have an isosurface of $\nu _{QPO} = 300Hz$ with a fixed
value of $n$, e.g., $n = 3$, and thus we can determine the minimum
and maximum values of $\xi _0 $, and obtain the corresponding values
of the parameter $n$. In this way we can fit the QPO frequencies of
several sources based on MCDD as given in Table.1.

\begin{figure}
\vspace{0.5cm}
\begin{center}
\includegraphics[width=6cm]{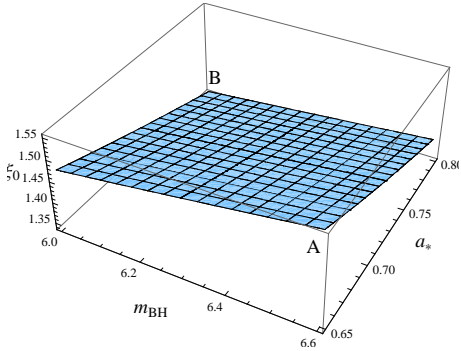}
\caption{An isosurface of $\nu _{QPO}=300Hz$ for GRO J1655-40 in the
parameter space consisting of the parameters $m_{BH},a_*$ and $\xi
_0 $. } \label{fig3}
\end{center}
\end{figure}

\begin{table*}
\caption{Fitting QPO frequencies of BH X-ray binaries based on MCDD.
}
\begin{tabular}
{|p{70pt}|p{62pt}|p{50pt}|p{46pt}|p{58pt}|p{52pt}|p{52pt}|} \hline
\raisebox{-3.00ex}[0cm][0cm]{\textbf{Source}}&
\multicolumn{3}{|p{158pt}|}{Input} &
\multicolumn{3}{|p{163pt}|}{Output}  \\
\cline{2-7}
 &
\raisebox{-1.50ex}[0cm][0cm]{$\nu _{QPO} $}&
\raisebox{-1.50ex}[0cm][0cm]{$m_{BH} $}&
\raisebox{-1.50ex}[0cm][0cm]{$a_\ast $}&
\raisebox{-1.50ex}[0cm][0cm]{$\xi _{\max } , \xi _{\min } $}&
$\lambda = 3$&
$n = 3$ \\
\cline{6-7}
 &
 &
 &
 &
 &
$n$&
$\lambda $ \\
\hline \raisebox{-4.50ex}[0cm][0cm]{GRO J1655-40}&
\raisebox{-1.50ex}[0cm][0cm]{300}& 6.6& 0.65& 1.554& 5.284&
1.547 \\
\cline{3-7}
 &
 &
6.0& 0.8& 1.338& 6.656&
1.343 \\
\cline{2-7}
 &
\raisebox{-1.50ex}[0cm][0cm]{450}& 6.6& 0.65& 2.287& 3.551&
2.366 \\
\cline{3-7}
 &
 &
6.0& 0.8& 1.628& 4.726&
1.649 \\
\hline \raisebox{-4.50ex}[0cm][0cm]{XTEJ1550-564}&
\raisebox{-1.50ex}[0cm][0cm]{184}& 10.8& 0.71& 1.485& 5.587&
1.483 \\
\cline{3-7}
 &
 &
8.4& 0.87& 1.233& 8.146&
1.242 \\
\cline{2-7}
 &
\raisebox{-1.50ex}[0cm][0cm]{276}& 10.8& 0.71& 2.036& 3.868&
2.083 \\
\cline{3-7}
 &
 &
8.4& 0.87& 1.397& 5.778&
1.418 \\
\hline \raisebox{-4.50ex}[0cm][0cm]{GRS 1915+105}&
\raisebox{-1.50ex}[0cm][0cm]{113}& 18.0& 0.98& 1.221& 7.122&
1.256 \\
\cline{3-7}
 &
 &
10.0& 0.998& 1.114& 12.044&
1.119 \\
\cline{2-7}
 &
\raisebox{-1.50ex}[0cm][0cm]{168}& 18.0& 0.98& 1.360& 5.294&
1.424 \\
\cline{3-7}
 &
 &
10.0& 0.998& 1.177& 8.714&
1.182 \\
\hline
\end{tabular}

\small
  Note. -- The values of the BH mass and spin of the sources are
adopted from Remilllard {\&} McClintock (2006), Davis et al. (2006)
and McClintock et al. (2006). The quantities $\xi _{\max } $ and
$\xi _{\min } $ are the maximum and minimum of $\xi _0 $ on the
isosurface of $\nu _{QPO} = const.$

\label{tab1}
\end{table*}

\subsection{Improving the fittings for 3:2 QPO pairs based on MCDD}

As is well known, the 3:2 QPO pairs have been observed in the BH
binaries listed in Table 1, i.e., GRO J1655-40, XTE J1550-564 and
GRS 1915+105 (MR06). Also, the 3:2 QPO pair appears in the BH
candidate H1743-322, and its behavior resembles the BH binaries XTE
J1550-564 and GRO J1655-40 in many ways (Homan et al. 2005; Kalemci
et al. 2006; Remillard et al. 2002, 2006).

Abramowicz {\&} Kluzniak (2001) explained the pairs as a resonance
between orbital and epicyclic motion of accreting matter. Recently,
the resonance model is presented in a more realistic context, in
which ``parametric resonance'' concept is introduced to describe the
oscillations rooted in fluid flow where there is a coupling between
the radial and polar coordinate frequencies (Abramowicz et al. 2003;
Kluzniak et al. 2004; T\"{o}r\"{o}k et al. 2005).

Not long ago, Aschenbach (2004, hereafter A04) found that the BH mass and
spin are strongly constrained by the 3:2 QPO pairs, and the ratio of the
vertical epicyclic frequency to the radial epicyclic frequency is given by

\begin{equation}
\label{eq10}
\left\{ {\begin{array}{l}
 \Omega _V (a_\ast ,\tilde {r}_{31} ) / \Omega _R (a_\ast ,\tilde {r}_{31} )
= 3, \\
 \Omega _V (a_\ast ,\tilde {r}_{32} ) / \Omega _R (a_\ast ,\tilde {r}_{32} )
= 3 \mathord{\left/ {\vphantom {3 2}} \right. \kern-\nulldelimiterspace} 2.
\\
 \end{array}} \right.
\end{equation}

In equation (\ref{eq10}) $\Omega _V $ and $\Omega _R $ are the
vertical and radial epicyclic frequencies, respectively (Nowak {\&}
Lehr 1998, Merloni et al. 1999). It has been found in A04 that
equation (\ref{eq10}) has only one solution for the two
commensurable orbits, i.e., $\tilde {r}_{31} = 1.546$ and $\tilde
{r}_{32} = \mbox{3.919}$ with an extremely high spin $a_ * =
\mbox{0.99616}$. As argued in A04, the radial gradient of the
orbital velocity of a test particle changes sign in a narrow annular
region when $a_
* > 0.9953$, and the BH mass of these binaries can be constrained in
a very narrow range, which are consistent with the dynamically determined
masses within their measurement uncertainty range.

It has been pointed out that QPOs are generally associated with the steep
power--law (SPL) state in BH X-ray binaries. Although the 3:2 QPO pairs
could be interpreted in some epicyclic resonance models, there remain
serious uncertainties as to whether epicyclic resonance could overcome the
severe damping forces and emit X--rays with sufficient amplitude and
coherence to produce the QPOs (e.g., see a review in MR06). The mechanism of
producing QPOs based on MCDD might remedy the disadvantage of the epicyclic
resonance models.

The 3:2 QPO pair has been observed in H1743-322 (Homan et al. 2005;
Remillard et al. 2006; Kalemci et al. 2006), and its behavior resembles the
BH binaries XTE J1550-564 and GRO J1655-40 in many ways (Remillard et al.
2002, 2006).

Based on the radial position of the vertical and radial epicyclic
resonance given by A04, we set $\tilde {r}_0 = \tilde {r}_{32} $ by
adjusting the parameters $n$ and $\lambda $ in equation (\ref{eq7})
with $a_
* = \mbox{0.99616}$. Thus we expect that both the vertical and
radial epicyclic resonance are stimulated and maintained due to
energy transferred magnetically from the inner footpoint to the
outer footpoint at $\tilde {r}_0 $. According to the strict
constraint required by the 3:2 QPO pair, we can estimate the BH mass
as well as $\nu _{MCDD} $ as listed in Table.2.

\begin{table*}
\caption{Fitting QPOs in BH binaries based on the epicyclic
resonance model. }
\begin{tabular}
{|p{72pt}|p{57pt}|p{64pt}|p{64pt}|p{64pt}|p{64pt}|} \hline
\raisebox{-1.50ex}[0cm][0cm]{\textbf{Source}}&
\raisebox{-1.50ex}[0cm][0cm]{${\nu _{up} } \mathord{\left/
{\vphantom {{\nu _{up} } {\nu _{down} }}} \right.
\kern-\nulldelimiterspace} {\nu _{down} }$}&
\raisebox{-1.50ex}[0cm][0cm]{$m_{BH} $}&
\raisebox{-1.50ex}[0cm][0cm]{$a_\ast , \quad \xi _0 $}&
\raisebox{-1.50ex}[0cm][0cm]{$\nu _{MCDD} $}&
\raisebox{-1.50ex}[0cm][0cm]{${\nu _{MCDD} } \mathord{\left/
{\vphantom {{\nu _{MCDD} } {\nu _{up} }}} \right. \kern-\nulldelimiterspace} {\nu _{up} }$} \\

 &
 &
 &
 &
 &
  \\
\hline \raisebox{-1.50ex}[0cm][0cm]{GRO J1655-40}&
\raisebox{-1.50ex}[0cm][0cm]{450 / 300}& 6.77&
\raisebox{-10.50ex}[0cm][0cm]{0.99616,  \par 2.9997}& 1371.51&
3.0478 \\
\cline{3-3} \cline{5-6}
 &
 &
6.75&
 &
1375.57&
3.0568 \\
\cline{1-3} \cline{5-6} \raisebox{-1.50ex}[0cm][0cm]{XTEJ1550-564}&
\raisebox{-1.50ex}[0cm][0cm]{276 / 184}& 11.06&
 &
839.521&
3.0417 \\
\cline{3-3} \cline{5-6}
 &
 &
11.02&
 &
842.568&
3.0528 \\
\cline{1-3} \cline{5-6} \raisebox{-1.50ex}[0cm][0cm]{GRS 1915+105}&
\raisebox{-1.50ex}[0cm][0cm]{168 / 113}& 18.49&
 &
502.169&
2.9891 \\
\cline{3-3} \cline{5-6}
 &
 &
17.77&
 &
522.516&
3.1102 \\
\cline{1-3} \cline{5-6} \raisebox{-1.50ex}[0cm][0cm]{H 1743--322}&
\raisebox{-1.50ex}[0cm][0cm]{240 / 160}& 12.70&
 &
731.111&
3.0463 \\
\cline{3-3} \cline{5-6}
 &
 &
12.66&
 &
733.421&
3.0559 \\
\hline
\end{tabular}
\\ \small Note. -- The BH masses of GRO J1655-40, XTEJ1550-564 and GRS
1915+105 are given by A04, and the the BH mass of H 1743--322 is
determined by the same way given in A04. The parameter $\xi _0 =
\mbox{2.9997}$ is determined by $\xi _0 \equiv {r_{32} }
\mathord{\left/ {\vphantom {{r_{32} } {r_{ms} }}} \right.
\kern-\nulldelimiterspace} {r_{ms} } = {r_0 } \mathord{\left/
{\vphantom {{r_0 } {r_{ms} }}} \right. \kern-\nulldelimiterspace}
{r_{ms} }$ for $a_\ast = \mbox{0.99616}$. \label{tab2}
\end{table*}

In the above fitting the parameters $n$ and $\lambda $ are not
independent, being related by equation (\ref{eq7}) for $a_\ast =
\mbox{0.99616}$ and $\tilde {r}_0 = \tilde {r}_{32} = \mbox{3.919}$.
And we have the curve of $n$ versus $\lambda $ given in Fig.4.
Inspecting Fig.4, we find that the power-law index $n$ increases
very sharply with the parameter $\lambda $, as the latter is small,
while $n$ increases very slowly with $\lambda $ as the latter is
great. These results imply that the fittings of QPOs based on MCDD
are very sensitive to the parameter $n$, if the radial width of
current distribution is small, while the fittings are almost
independent of $n$, provided that the current distribution is wide
enough.

\begin{figure}
\vspace{0.5cm}
\begin{center}
\includegraphics[width=6cm]{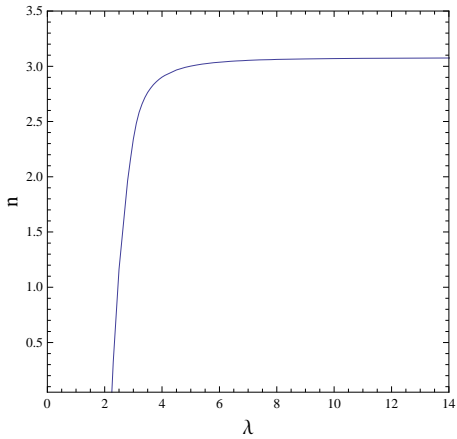}
\caption{The curve of the parameters $n$ versus $\lambda $
constrained by the 3:2 QPO pairs of the BH binaries.}
 \label{fig4}
\end{center}
\end{figure}

It is interesting to note that $\nu _{MCDD} $ for each source in
Table 2. is about three times the corresponding upper frequency of
the 3:2 QPO pair, i.e., $\nu _{MCDD} \approx 3\nu _{up} $. Based on
general relativity we have ${\nu _{MCDD} } \mathord{\left/
{\vphantom {{\nu _{MCDD} } {\nu _{up} }}} \right.
\kern-\nulldelimiterspace} {\nu _{up} } = 3.05$, which is
independent of the BH mass. This result seems consistent with the
commensurate frequencies of the 3:2 QPO pair. In addition, the
combination of MCDD with the epicyclic resonance might be helpful to
interpret the association of the QPOs with the SPL states in BH
binaries, because both the steep power--law component of radiation
and the QPOs in BH binaries have the same origin of the magnetic
reconnection. We shall address this issue in details in our future
work.

\noindent\textbf{Acknowledgments. } This work is supported by the
National Natural Science Foundation of China under grant 10873005,
the Research Fund for the Doctoral Program of Higher Education under
grant 200804870050 and National Basic Research Program of China
under grant 2009CB824800.

\end{document}